\documentclass[conference]{IEEEtran}
\usepackage{algorithm,algorithmic,amsfonts,graphicx,enumerate,amsmath,pst-all,comment,balance}

\begin{document}
\bibliographystyle{IEEEtran}

\title{Using a Market Economy to Provision Compute Resources Across Planet-wide Clusters}

% BHY: These should be placed in order of contribution. I'll leave that to others. :-)
% MS: I made a first attempt, though it is not clear to me that it is the correct one.
\author{
\IEEEauthorblockN{Murray Stokely, Jim Winget,\\
Ed Keyes, Carrie Grimes}
\IEEEauthorblockA{Google, Inc.\\
Mountain View, CA\\
\{mstokely,winget,edkeyes,cgrimes\}@google.com}
\and
\IEEEauthorblockN{Benjamin Yolken}
\IEEEauthorblockA{Department of Management Science and Engineering\\
Stanford University\\
Stanford, CA\\
yolken@stanford.edu}
}

\maketitle

\begin{abstract}

We present a practical, market-based solution to the resource
provisioning problem in a set of heterogeneous resource clusters. We
focus on provisioning rather than immediate scheduling decisions to
allow users to change long-term job specifications based on market
feedback. Users enter bids to purchase \emph{quotas}, or bundles of resources for
long-term use. These requests are mapped into a simulated \emph{clock
  auction} which determines uniform, fair resource prices that balance
supply and demand. The reserve prices for resources sold by the
operator in this auction are set based on current utilization, thus
guiding the users as they set their bids towards under-utilized
resources. By running these auctions at regular time intervals, prices
fluctuate like those in a real-world economy and provide motivation
for users to engineer systems that can best take advantage of
available resources.

These ideas were implemented in an experimental resource market at
Google. Our preliminary results demonstrate an efficient transition
of users from more congested resource pools to less congested
resources. The disparate engineering costs for users to reconfigure
their jobs to run on less expensive resource pools was evidenced by
the large price premiums some users were willing to pay for more
expensive resources. The final resource allocations illustrated how
this framework can lead to significant, beneficial changes in user
behavior, reducing the excessive shortages and surpluses of more
traditional allocation methods.

\end{abstract}

%\begin{keywords}grid economy; market-based systems; world-wide computing; resource
%  allocation\end{keywords}

\section{Introduction}
The last decade has seen the rise of IT services supported by large, heterogeneous, distributed computing systems. These systems, which serve as the backbone for ``grid,'' ``cloud,'' and ``utility'' computing environments, provide the necessary speed, redundancy, and cost-effectiveness to enable the powerful Internet applications offered by Google and others. In addition, such systems can be used internally, within the enterprise, to support essential business functions. In the future, as costs per unit continue to go down, we will see exciting new functionality and applications supported by these environments.

Despite their size, however, the resources in these large-scale, distributed systems are necessarily finite. Not every external client or internal engineering team can get everything that it demands. The system operator must  place hard limits on the CPU, disk, memory, etc. that each job or job class can use. Otherwise, some tasks will be starved and user experience will suffer. These allocation limits are then mapped into the low-level scheduling algorithms (e.g., time sharing) used to actually assign jobs to units of physical hardware.

Traditionally, such limits / quotas have been set manually according to pre-defined policies. The operator either grants each user an equal share of the system or, more likely, decides that certain jobs / users are ``more important'' than others, giving the former higher quotas or the ability to preempt lower-ranked tasks. This approach works well for small, homogeneous systems, but scales poorly as the systems and their supported user bases get larger and more heterogeneous. In the latter case, the information requirements on the operator can become extremely onerous; even with complex optimization procedures, the final allocation is often \emph{inefficient} from a system-wide perspective. These inefficiencies are manifested though uneven utilization, significant shortages and surpluses in certain resource pools, and general user unhappiness.

One solution is to instead allocate resources through a \emph{market-based} system. Indeed, this is the manner in which nearly all commodities (e.g., oil, food, real estate, etc.) are priced and allocated in the ``real-world'' economy. People use money to purchase goods and services in such a way as to maximize their individual happiness. Participants with excess resources, on the other hand, sell them to receive money. The prices for these goods and services, in turn, dynamically fluctuate so as to match supply and demand. If the latter forces are not well-understood, then one can use \emph{auctions} to gauge these and set prices appropriately. In this way, goods are allocated efficiently without quotas or extensive centralized control.

Much literature has proposed applying these market principles to the
large-scale computing systems under study. Many of these models use
auction-based pricing for the scheduling of \emph{individual jobs}
across multiple administrative domains
\cite{Buyya02economicmodels,689702,10.1109/IPDPS.2001.924985}. Within
this environment, these authors propose replacing standard scheduling
algorithms with ``economically inspired'' ones. Round robin, for
instance, might be replaced by a simulated auction in which the
priority of each job is mapped into a ``bid'' \cite{Lai,121753}.

While the previous models are potentially useful for low-level
scheduling, our focus here is on the higher-level provisioning of
\emph{aggregate resources} (e.g. quotas) across users within a single
administrative domain that spans sites around the world. These types of allocation problems have been studied extensively from a theoretical standpoint, but few practical applications have been introduced and evaluated. Those few appearing in the literature have been  small-scale and limited to very specific application domains \cite{Regev,779943}. As discussed in \cite{Auyoung07practicalmarket-based}, much work remains to be done in making these ``market-based'' approaches both practical and scalable.

In this paper, we propose a novel, high-level approach to the pricing
and allocation of resources in large-scale, ``grid'' systems. In
particular, users specify desired bundles of resources along with the
maximum amount they are willing to pay (or, if selling, the minimum
amount they are willing to receive). In addition, the system operator
acts as a seller of resources and sets
\emph{reserve prices} based on the current utilization of each
resource pool. The final winners and prices are then determined
through a \emph{clock auction} starting at the latter prices. By
repeating this process on a regular time scale (e.g., weekly), one can
thus develop a dynamic, efficient ``economy'' for computing resources
within a distributed, heterogeneous environment.

Given this framework, we implemented a combinatorial exchange within Google and ran experiments in which this was used to allocate resources across the company's engineering teams. Our preliminary results indicate that this approach offers several advantages over traditional allocation mechanisms:

\begin{enumerate}
\item Allows market participants to make engineering design tradeoffs
  with the jobs they choose to run on their allocated resources.
  Teams that find resource A at a significant discount to resource B
  may bid on resource A and set about reengineering their job to use
  less of resource B and more of resource A.

\item Long-term resource utilization can be taken into account when setting
  reserve prices for new resources brought into the marketplace. In
  this way bidding behavior can be encouraged that improves the overall
  bin-packing of system clusters.  Specifically, in later sections
  we develop a utilization weighted reserve price calculation to
  encourage uniform utilization levels across resource pools.
\end{enumerate}

The remainder of this paper is organized as follows. We begin in the next section by discussing our mathematical model for resources and user preferences in a large-scale computing environment. Section 3 describes the clock auction mechanism used to set prices in our system. Section 4 addresses the utilization-based reserve pricing scheme described above. Our experimental results are discussed in Section 5. Finally, in Section 6, we conclude and give directions for future research.

\section{Resource and User Preference Models}
Consider a large-scale computing system with $R$ resource pools, indexed as $r = 1, 2, \ldots, R$ and $U$ users, indexed as $u = 1, 2, \ldots U$. The former represent aggregations of individual, physical resources which are divisible and shared across the users. Although the units of resources are the same for all resource pools, the pools in practice need not have exactly the same characteristics. Geographic location, location of other required resources or data, network connectivity, or other secondary characteristics may (or may not) distinguish a particular pool for a particular user. In the experiments described later, these pools represented high-level, resource / location pairs (e.g., ``CPUs in cluster 1''). The exact criteria used to distinguish these groups, however, is flexible and ultimately depends on the system being modeled. Similarly, the definition of a ``user'' is also flexible; in the case of our experiments, these were taken as distinct engineering teams within the company.

These users, given some initial endowment of money and resources, seek to buy, sell, and trade the latter in an individually optimal way. The exact details of this optimization are application-dependent. In many ``grid'' settings, however, the underlying user preferences are highly \emph{combinatorial}; that is, users get maximal utility from particular bundles of resources consisting of very specific resource combinations. CPUs in a particular place, for instance, are probably not useful unless the user can get colocated memory, disk, and network resources as well. At the same time, these preferences may also reveal strong substitutability among resource bundles. For example, a user may demand a certain combination of CPU, memory, and disk but may be indifferent with respect to the exact location.

These preferences are well-understood by the users but not necessarily by the system operator. As discussed in the introduction, therefore, we
propose that prices and allocations be set through an
\emph{auction}. In particular, users announce bids encapsulating their
desired bundles and ``willingness to pay'' criteria in a tree-based bidding
language similar to \emph{TBBL} described in \cite{1064036}. More formally, we assume that each user $u$ submits some bid, $\mathcal{B}_u$, consisting of the pair $\{\mathcal{Q}_u, \pi_u\}$. The first term is a set of R-component vectors, $q_u^1, q_u^2, \ldots$ representing bundles of resources over which $u$ is indifferent. In other words, $u$ desires
[$q_u^1$ XOR $q_u^2$ XOR $q_u^3$ $\ldots$]. The components of each
vector, in turn, encode either the quantities of each resource desired
(if positive) or offered (if negative). The second bid term, $\pi_u$,
gives the total, maximum amount that $u$ is willing to pay (if
positive) or the minimum amount that $u$ is willing to receive (if
negative). We assume that this is a scalar value applying to all bundles in $\mathcal{Q}_u$. Extending the model to allow for vector $\pi$'s, corresponding to distinct valuations for each individual user bundle, does not significantly change our results and is omitted for notational simplicity.

We also assume that all bids are entered simultaneously, i.e. users
cannot modify their bids based on the actions of others or provisional
feedback from the system. In practice, participants are typically
given some window of time in which to enter bids and, possibly,
respond to environmental conditions. In this case, our analysis here
still holds if we instead consider the bids to be the \emph{final}
values as of the end of the entry period. Modeling the dynamics of
this initial phase is a very interesting problem, but one that is outside the scope of our investigation here.
%this initial phase is a very interesting problem but has not yet been
%studied and hence is left as a topic for future research.

\section{Pricing Algorithm}
\subsection{Design Goals}
Given the auction setup above, we now seek some algorithm that maps the bids described previously into both resource \emph{prices} and \emph{allocations}. The former are important for two reasons. First, these determine how much each winning user pays for its awarded resources. Second, and perhaps more importantly, though, these final prices act as \emph{signals} to both the operator and the users. If demand exceeds supply for some resource pool, for instance, then this should be reflected by a significant price increase. This increase, in turn, indicates to losing bidders that they should increase their bids in future auctions. Moreover, it indicates to the system operator that there may be a shortage in the corresponding pool; the operator should address this shortage by increasing the supply of resources appropriately.

Any pricing algorithm, therefore, should satisfy the following criteria:

\begin{enumerate}
\item \textbf{Clear Signaling:} As discussed above, one of the primary goals of a grid-system resource auction is to provide clear signals about supply and demand to the participants. This is easiest to achieve when the final prices are \emph{uniform} and \emph{linear}. In other words, all winners within a given resource pool pay exactly the same ``price per unit.'' These unit prices are clearly announced at the end of the auction so that losing bidders can adjust their behavior in future auctions.

\item \textbf{Fair Outcomes:} Winners should be those who have ``bid enough'' and losers those who have bid ``too little'' given the final auction prices. In this sense, allocations are determined completely by the prices and bids, not by, ``unfair,'' exogenously provided operator preferences.

\item \textbf{Computational Tractability:} In a large grid computing environment, any pricing and allocation method has to scale well in the size of the system (i.e., number of resource pools and users). Procedures that require solving NP-hard optimizations are probably not a good choice.

\end{enumerate}

\subsection{Mathematical Description}
We now describe the problem in more formal, mathematical terms. Let the final auction prices be given by the R-component vector $p$ and let $x_1, x_2, \ldots, x_U$ represent the corresponding user allocations. Furthermore, let $\mathbb{W}$ and $\mathbb{L}$ represent, respectively, the sets of ``winning'' and ``losing'' bidders. Our algorithm design problem can then be expressed as

\vspace{5mm}

\noindent SYSTEM:
\begin{equation*}
\begin{array}{rccll}
\mathop{\textrm{max}}_{x,p} & \multicolumn{2}{c}{f(x,p)} & & \\
\textrm{subject to:} & x_u & \in & \{0 \cup \mathcal{Q}_u\} & \forall u \\
& \sum_u x_u & \leq & 0 &\\
& \pi_u & \geq & x_u^{\textrm{T}}p & \forall u \in \mathbb{W}\\
& x_u^{\textrm{T}}p & = & \mathop{\textrm{min}}_{q \in
\mathcal{Q}_u} q^{\textrm{T}}p & \forall u \in \mathbb{W}\\
& \pi_u & < & \mathop{\textrm{min}}_{q \in \mathcal{Q}_u}
q^{\textrm{T}}p & \forall u \in \mathbb{L}\\
& p & \geq & 0 & \\
\end{array}
\end{equation*}

\noindent We refer to the above optimization as SYSTEM in the sequel.

There are many possible forms for the objective function, $f(x,p)$, and the ``best'' choice from these varies significantly from application to application. Some possibilities include using (1) the total \emph{surplus}, i.e. the total difference between what users are willing to pay minus what they actually pay or (2) the total \emph{value of trade}, i.e. the net value of all resources that change hands. Optimizing over these and then comparing the relative costs and benefits of the outcomes produced are rich areas for future research. In this paper we focus on a tractable solution for satisfying the constraints. Note that these have the following interpretations:

\begin{itemize}
\item[(1)]
$x_u \in \{0 \cup \mathcal{Q}_u\} \ \forall u$: Users either get one
of their desired bundles or nothing at all. Thus, no scaling is
allowed.
\item[(2)]
$\sum_u x_u \leq 0$: The final allocation leads to a net surplus of resources. There are no shortages created by awarding resources which are not available for distribution.
\item[(3)]
$\pi_u \geq x_u^{\textrm{T}}p\ \forall u \in \mathbb{W}$: All winners
have provided values that are ``high enough'' given final prices.
\item[(4)]
$x_u^{\textrm{T}}p = \mathop{\textrm{min}}_{q \in \mathcal{Q}_u}
q^{\textrm{T}}p\ \forall u \in \mathbb{W}$: Winners attain the
``cheapest'' bundles in their indifference sets.
\item[(5)]
$\pi_u < \mathop{\textrm{min}}_{q \in \mathcal{Q}_u} q^{\textrm{T}}p\
\forall u \in \mathbb{L}$: All losers have bid ``too low'' given the
final prices.
\item[(6)]
$p \geq 0$: Prices must be non-negative.
\end{itemize}

Alternatively, one can tighten the second constraint to $\sum_u x_u =
0$, i.e. eliminate all ``left over,'' surplus resources, at the expense of
violating the first one. Note that, in practice, it is usually impossible to have the market perfectly clear without scaling some user; supplies and demands will rarely ``align'' to the necessary extent.

Another possible modification is to replace the last constraint by $p
\geq p_{\textrm{min}},\ p \leq p_{\textrm{max}}$ where
$p_{\textrm{min}}$ and $p_{\textrm{max}}$ are reasonable lower and
upper bounds, respectively, on system prices. Doing so can keep the
system away from ``weird'' or ``unfair'' values. On the downside,
though, these additional constraints reduce the size of the feasible
region and increase the possibility that no solution exists. In the
remainder of this document, we assume for simplicity that $p$ is only
constrained to be nonnegative. Replacing this with upper and lower
bounds does not significantly change the analysis or our results.

Finally, one might want to relax the win / loss constraints in the
case of \emph{ties}. For instance, if there is one unit of supply, and
two users are both willing to pay up to \$1.00 for this unit, then the
only feasible, ``fair'' outcome is that both lose and the resource
remains unallocated. As this seems wasteful, it may be desirable in
this case to break the ``fairness'' constraints by allowing one to
lose and the other to win. This is less of a worry in large systems,
though, because the probability of having an exact tie is extremely
low.

\subsection{Ascending Clock Auction}\label{sec:clockauction}
The requirements above rule out the most commonly applied combinatorial auction algorithms (e.g., the Vickrey-Clarke-Groves (VCG) mechanism), as these are generally \emph{not} computationally tractable and do not produce fair, uniform prices without significant ``post processing'' (see \cite{Cramton20062} for a more detailed discussion).

Instead, we propose using an \emph{ascending clock auction} inspired by the ``Simultaneous Clock Auction'' from \cite{Ausubel2004} and the updated, more advanced ``Clock-Proxy Auction'' from \cite{Cramton2006}. In both of these, the current price for each resource is represented by a ``clock'' that all participants can see. At each time slot, users name the quantities of each resource desired / offered at the given price. If the excess demand vector, $\sum_u x_u$, has all nonpositive components, then the auction ends. Otherwise, the auctioneer increases the prices for those items which have positive excess demand, and again solicits bids. Hence, the auction allows users to ``discover'' prices in real time and ends with a ``fair'' allocation in which all users pay / receive payment in proportion to uniform resource prices.

This ``clock'' approach has many desirable properties (detailed later) but requires feedback from the users over multiple rounds. We can adapt the algorithm to our single-round environment, however, by introducing \emph{bidder proxies} that automatically adjust their demands on behalf of the real bidders. Although these proxies can be extremely sophisticated in practice, for our purposes here we model them as functions, $\mathcal{G}_u(p) \rightarrow q_u$, mapping the current prices into the ``best'' bundle from each $\mathcal{Q}_u$ set. In particular, we have

\begin{eqnarray}
 \mathcal{G}_u(p) & = & \left\{ \begin{array}{ll} \hat q_u &
    \textrm{if } \hat q_u^{\textrm{T}} p \leq \pi_u \\ 0 &
    \textrm{otherwise}\end{array}\right. \\
\textrm{where} ~~~ \hat q_u & \in & \mathop{\textrm{arg min}} q_u^{\textrm{T}} p
\end{eqnarray}

%\noindent where
%\begin{equation}
%\end{equation}

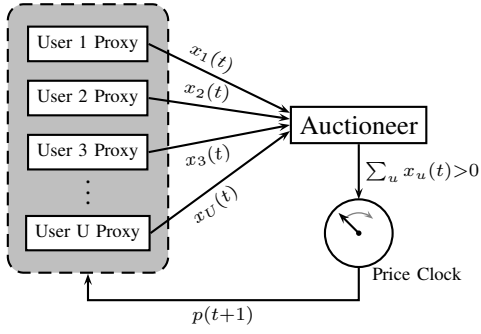
\begin{figure}[t]
\centering \psset{unit=1.8mm}
\begin{pspicture}(25,25)

\psframe[fillstyle=solid,fillcolor=lightgray,framearc=0.2,linestyle=dashed](-4,5)(8,25)

\rput(2,22){\rnode{U1}{\psframebox[fillstyle=solid,fillcolor=white]{$\scriptstyle{\textrm{User 1 Proxy}}$}}}
\rput(2,18){\rnode{U2}{\psframebox[fillstyle=solid,fillcolor=white]{$\scriptstyle{\textrm{User 2 Proxy}}$}}}
\rput(2,14){\rnode{U3}{\psframebox[fillstyle=solid,fillcolor=white]{$\scriptstyle{\textrm{User 3 Proxy}}$}}}

\pnode(2,5){UN}

\rput(2,11.6){$\vdots$}

\rput(2,8){\rnode{UU}{\psframebox[fillstyle=solid,fillcolor=white]{$\scriptstyle{\textrm{User U Proxy}}$}}}

\rput(22,16){\rnode{A}{\psframebox{Auctioneer}}}
\rput(22,8){\Cnode[radius=2.5]{C}{}}

\rput(26.7,12.5){$\scriptstyle{\sum_u x_u(t) > 0}$}
\rput(26.3,5){$\scriptstyle{\textrm{Price Clock}}$}

\pscircle*(22,8){0.2}
\psarc[linewidth=0.5pt,linecolor=gray]{<-}(22,8){1.5}{45}{135}
\psline{->}(22,8)(20.5,9.5)

\ncdiag[angleA=0,angleB=170,arm=0]{->}{U1}{A}
\naput[nrot=:U,labelsep=0.3,npos=1.4]{$\scriptstyle{x_1(t)}$}

\ncdiag[angleA=0,angleB=175,arm=0]{->}{U2}{A}
\naput[nrot=:U,labelsep=0.3,npos=1.4]{$\scriptstyle{x_2(t)}$}

\ncdiag[angleA=0,angleB=180,arm=0]{->}{U3}{A}
\nbput[nrot=:U,labelsep=0.3,npos=1.4]{$\scriptstyle{x_3(t)}$}

\ncdiag[angleA=0,angleB=-175,arm=0]{->}{UU}{A}
\nbput[nrot=:U,labelsep=0.3,npos=1.4]{$\scriptstyle{x_U(t)}$}

\ncline{->}{A}{C}

\ncbar[angle=-90]{->}{C}{UN}
\naput[labelsep=0.3]{$\scriptstyle{p(t+1)}$}

\end{pspicture}
\caption{Schematic of price update loop in ascending clock auction. The auctioneer collects the demands and/or supplies from each bidder proxy as a function of the current price. If demand exceeds supply, then the price clock is increased and the process repeats.}
\label{fig:clockprogression}
\end{figure}

Given this proxy framework, we can time index the system state and
then run a simulated, multi-round clock auction, as described below
and shown in Figure \ref{fig:clockprogression}.

\subsubsection{Algorithm}\label{sec:clockalgorithm}
From the discussion above, we have the following settlement algorithm:

\vspace{3cm}

\begin{algorithm}[h!]
\caption{Ascending Clock Auction}
\begin{algorithmic}[1]
\STATE Given: $U$ users, $R$ resources, starting prices $\tilde p$, update increment function $g: (x,p) \rightarrow \mathbb{R}^{R}$
\STATE Set $t = 0$, $p(0) = \tilde p$
\LOOP
\STATE Collect bids: $x_u(t) = \mathcal{G}_u(p(t))\ \forall u$
\STATE Calculate excess demand: $z(t) = \sum_u  x_u(t)$
\IF{$z(t) \leq 0$}
\STATE Break
\ELSE
\STATE Update prices: $p(t+1) = p(t) + g(x(t),p(t))$
\STATE $t \leftarrow t+1$
\ENDIF
\ENDLOOP
\end{algorithmic}
\end{algorithm}

\subsubsection{Parameters}\label{sec:clockparams}
As formulated above, the clock auction requires two input parameters: (1) $\tilde p$, a set of starting prices, and (2) $g(x,p)$, a function for setting the price increment. The first should be set well below the expected settling prices (to ensure positive excess demand). On the other hand, setting these too low may unnecessarily prolong the auction. If there exist ``reserve'' prices that serve as lower bounds on all bids, then these are a reasonable choice. See Section \ref{sec:reservepricing} below for one possible method of setting the former.

The $g(\cdot)$ function, on the other hand, maps the current system state into an R-dimensional vector of nonnegative, additive price updates. The simplest choice is $g(x,p) = \alpha z(t)^{+}$ where $\alpha$ is some small positive scalar, $z(t)$ is the excess demand defined above, and the notation $x^{+}$ is equivalent to $\mathop{\textrm{max}}(x,0)$, taken componentwise. In practice, however, this often causes the prices to move too quickly in the early rounds of the auction and then too slowly in the later ones. A more effective choice is to construct $g$ such that no price changes by more than some fixed fraction, say $\delta$:

\begin{equation}\label{eq:maxincrement}
g(x,p) = \mathop{\textrm{min}} (\alpha z(t)^{+}, \delta e)
\end{equation}

\noindent where $e$ is the R-dimensional vector of all $1$'s and the latter minimization is taken componentwise.

Another adjustment to consider for $g(\cdot)$ is a normalization for differences in the base resource prices. If some resource (e.g. disk) is much cheaper per unit than the others (e.g., CPU and RAM), then it may be helpful to reduce its increment rate. Otherwise, the final prices may be out of proportion from their expected relative sizes.

\subsubsection{Convergence}
Even if there exists a solution to SYSTEM, there may exist price directions along which excess demand is never eliminated. Hence, it is plausible that the clock auction may not converge even though its prices increase monotonically from round to round. We note, however, that such convergence is guaranteed if all participants are either ``pure buyers'' or ``pure sellers,'' i.e. each element of $\mathcal{Q}_u$ is either all nonnegative or all nonpositive for each user. The reasoning behind this is that, for each ``pure buyer'' there exists some price threshold at which this user ``drops out'' and no longer desires anything. Taking the maximum across all buyers gives us a ``price ceiling'' beyond which the auction must end.

When ``traders'' are also present, the convergence issue becomes much more complex. In fact, there are relatively small counterexamples with these types of users in which the clock auction never converges. These, however, are rather contrived and unlikely to be encountered in practice. Moreover, we hypothesize that there exist simple conditions (e.g., ``traders'' seek to buy more than they sell) which prevent these types of results. Proving these formally, however, is quite tricky and left as a topic for future research.

\subsubsection{Discussion}
Despite its apparent simplicity, the clock auction has a number of highly desirable features. First, and most significantly, it is computationally tractable. All else being equal, the execution time scales linearly in the number of participants and the number of resources. Solving for the prices in our experimental resource auction (having around $100$ bidders and $100$ system-level resources), for instance, took only a few minutes despite the fact that the underlying code was written in Python and was highly non-optimized. Optimized code written in a lower-level language could reduce this by at least one order of magnitude. See Section \ref{sec:results} below for more details.

Second, it is extremely robust. Irrespective of the system size, trade-dependencies between resources, the exact starting point, and other parameters, we have observed that it quickly reaches a ``reasonable'' set of prices and allocations. Unlike other algorithms, whose performance is extremely sensitive to the initial system parameters, the clock auction consistently generates plausible results on the first try.

Third, provided that it converges, the clock auction necessarily finds a feasible point for SYSTEM, i.e. a $(x,p)$ pair at which there is no excess demand, prices are uniform, allocations are ``fair,'' etc. Just getting this is a very hard problem in larger systems, particularly when a significant fraction of participants are traders.

The obvious downside, however, is that this clock procedure does not explicitly optimize anything; while it finds a feasible point, it completely ignores the objective function, $f(x,p)$, in its update steps. If there are multiple feasible points for SYSTEM (as there usually are), then there is no guarantee that the procedure will converge to the optimal one or even to one that is ``near optimal.'' This suggests that other algorithms, based explicitly on optimization, may be better choices if they can be implemented in a computationally tractable way. This is an exciting area of research, and one that we will detail in a future paper.

\section{Congestion-weighted Reserve Pricing}\label{sec:reservepricing}

The performance of the clock auction depends heavily on the starting /
reserve prices, $\tilde p$, chosen by the operator. As discussed
above, these serve to guide the users as they calculate their bids and
also affect the convergence speed of the clock procedure. More
broadly, however, the reserve prices form a key input related to the
\emph{economic engineering} of the market for compute resources
\cite{976142}. Specifically, the reserve prices form the basis of a
decision support framework in the market economy that allows the
operator to steer the system towards particular, desired outcomes. If
one resource pool is particularly crowded, for instance, then the
operator can set its reserve price high to ensure that users in this
pool have the incentive to leave it for another, less crowded
one. This also ensures that the operator gets a ``fair price'' when
bringing new resources online. Ideally, the market handles these
things automatically; this priming, however, may be necessary as the
economy is ``started up,'' when participation is low and/or users do
not fully understand how to optimally set their bids.  In such cases
of limited liquidity the use of a reserve pricing strategy based on
some operator utility function allows the market to degrade gracefully into a market with fixed but differential pricing set to influence bidder behavior in the desired way.

As in \cite{4536241} and \cite{383620}, we use an approach that takes into account the resource loads. For each resource pool $r$, we assume
there is a well-defined metric of current (i.e., pre-auction)
utilization, $\psi(r)$. Furthermore, assume that each resource pool, $r$, has a real, known cost $c(r)$. We then define our weighted reserve price for one unit of $r$ as:

\begin{equation}
\tilde p_r = \phi_r(\psi(r)) c(r)
\end{equation}

\noindent where $\phi_r(\cdot)$ is the weighting function for $r$.

\begin{figure}[t]
\centering
\includegraphics[width=8cm]{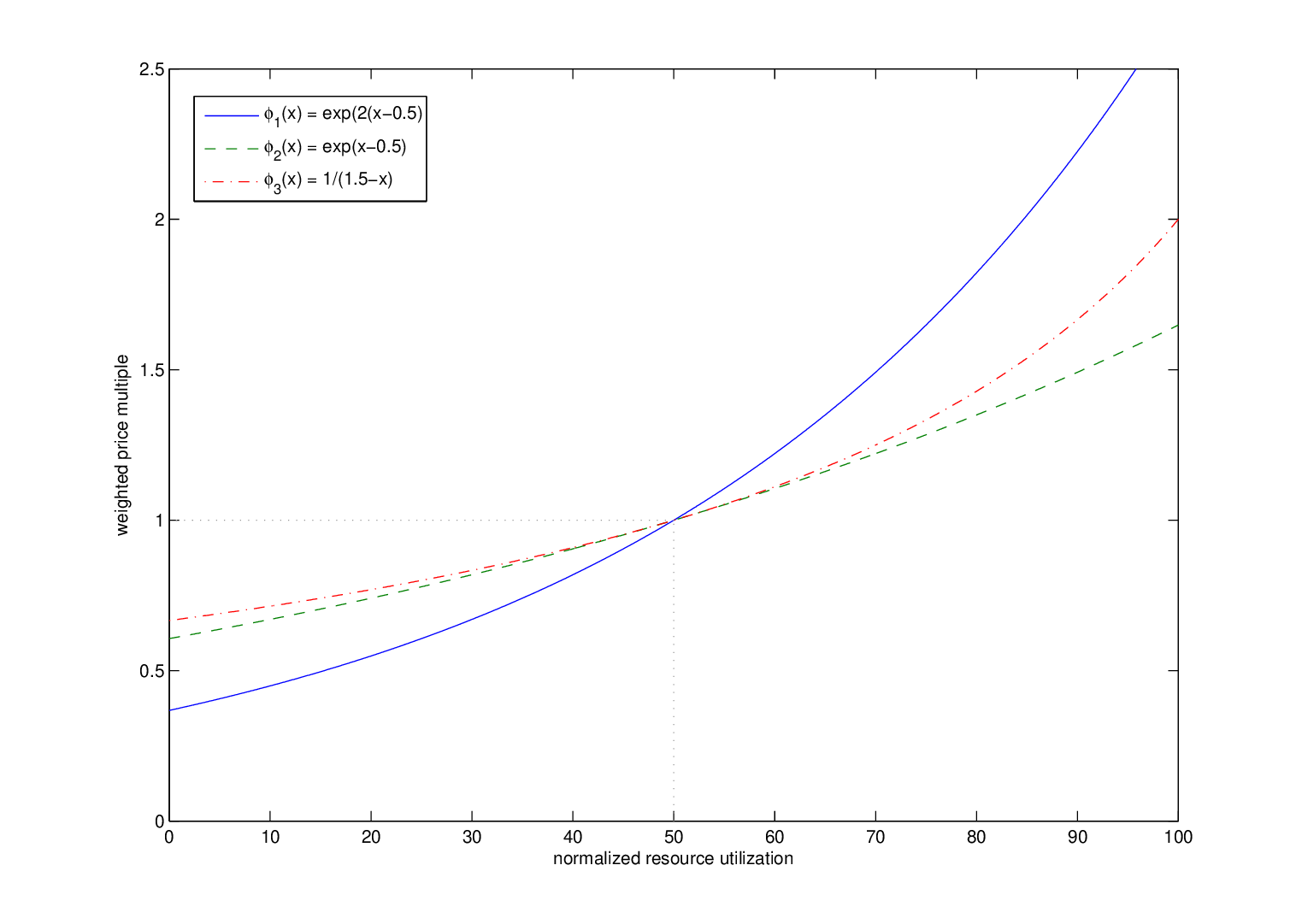}
\caption{Example utilization-weighted pricing curves.}
\label{fig:pricecurves}
\end{figure}

\subsection{Weighting Function Properties}
The weighting functions should accurately reflect the scarcity or
abundance of resources in individual pools. In particular, we propose
the following criteria for constructing these:

\begin{enumerate}
\item $\phi_r(\cdot)$ is monotonically increasing
\item $\phi_r(\cdot) > 1.0$ for resources that are overutilized
  in a cluster
\item $\phi_r(\cdot) \leq 1.0$ for resources that are underutilized
  in a cluster
\item The relative cost difference of resources in highly congested
  clusters (e.g. 99\% vs 80\% utilization) is significantly
  greater than the cost difference of resources in underutilized
  clusters (e.g. 40\% vs 15\% utilization) as the operator does
  not seek to encourage moves between underutilized clusters
\item $\phi_r(100\% \textrm{ utilization}) = k\phi_r(0\% \textrm{
  utilization})$, for some constant $k$, to limit the impact on the
  initial endowment of budget dollars
\end{enumerate}

The inputs of the weighting functions are utilization percentiles for
the different resource dimensions (e.g. CPU, RAM, disk, network). The
fifth property is strongly related to the strategy used for
disbursement of initial budget dollars among bidders and is not
covered here. Figure \ref{fig:pricecurves} shows some example weighting functions that were used in our market experiments.

\begin{comment}
\subsection{Reserve Price Effects on Money Supply}

Each user $u$ has a budget $B_u$ and seeks to procure a bundle of
resources $Q_u = {q^1_u, q^2_u, ...}$. These users can be expected to
take advantage of market pricing to choose clusters where the
resources they need are relatively cheap, as long as those clusters
meet other user-specific constraints about data-locality, network
connectivity, and service dependencies. We refer to the new bundle of
resources requested in response to changes in resource pricing as
$\tilde{Q_u} = {\tilde{q}^1_u, \tilde{q}^2_u, ...}$.

Given a model of user behavior $M$, budgets $B_u \forall u \in U$, and
a weighted pricing function $\phi_r(\cdot)$, we have found it useful
to simulate the difference in the final settled prices of the clock
auction between $\tilde{Q_u}$ and $Q_u$ $\forall u \in U$. The sum of
these price differences for each $u \in U$ can provide information
about how the team budgets could be scaled to take into account the
opportunities available in the market. The development of models of
user behavior is heavily dependent on the user needs and
substitutability of different resource pools, and so is not discussed
in a general context here.
\end{comment}

\section{Experimental Results}\label{sec:results}
Building a market economy inside a commercial entity requires an
extensive \emph{commercialization stack} separate from the distributed
applications and cluster management systems that actually use these
resources \cite{976142}.  In particular, the ideas from the previous
sections were used as the basis for decision support, price
formulation, and trading support layers of an experimental resource
allocation economy within Google. Other important systems for
accounting, billing, and contract management are not discussed here as
they are outside the scope of this paper.

To build the trading system, each resource pool was taken as a cluster / resource type combination with the latter including CPU, RAM, and disk.
Engineering teams were given budget dollars and allowed to buy, sell,
and trade resources with each other as well as the company
itself. These transactions were priced by running periodic clock
auctions with utilization-weighted reserve prices of the type
described above.

In the remainder of this section, we discuss the various pieces of this system and then evaluate the pricing and provisioning outcomes produced.

\subsection{The Trading Platform Implementation}
\begin{figure}[t]
\centering
\includegraphics[width=8cm]{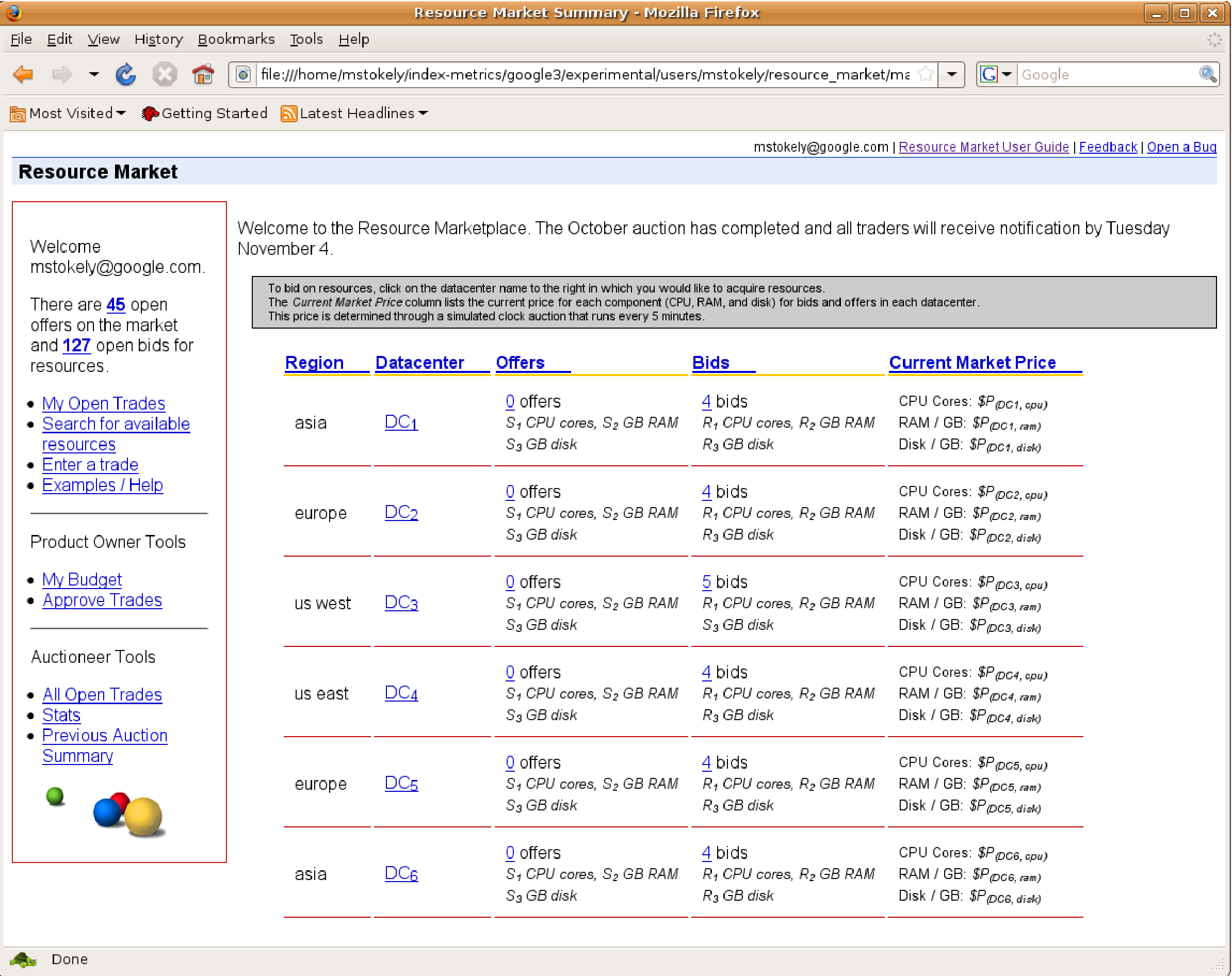}
\caption{Template for ``market summary'' page on trading platform front end. }
\label{fig:rmsnapshot}
\end{figure}

The trading platform front end was implemented as an internal web application
that could be used by teams to express trades in a simple bidding
language against the available clusters participating in the
auction.  Users are greeted on a main page ``market summary'' (see
Figure \ref{fig:rmsnapshot}) that lists the participating clusters
along with the number of active bids and offers in each, and the
current market prices as determined by the clock auction.

\begin{figure}[t]
\centering
\includegraphics[width=8cm]{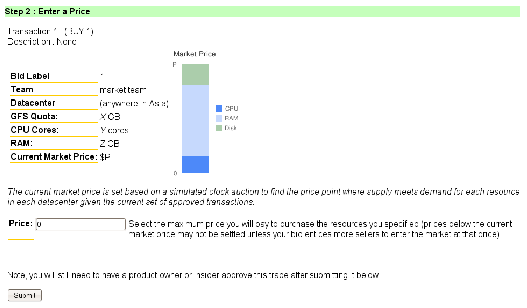}
\caption{Template for bid entry page. }
\label{fig:rmbid}
\end{figure}

Market participants enter bids through a two step process where
they first enter requirements in terms of desired cluster resources
(such as GFS \cite{945450} or Bigtable \cite{1298475} resources).  The
second step, as shown in Figure \ref{fig:rmbid}, displays the covering
amount of CPU, RAM, and disk and the current market prices for those
components before allowing the user to enter a maximum bid price (for
bids) or a minimum reserve price (for offers).

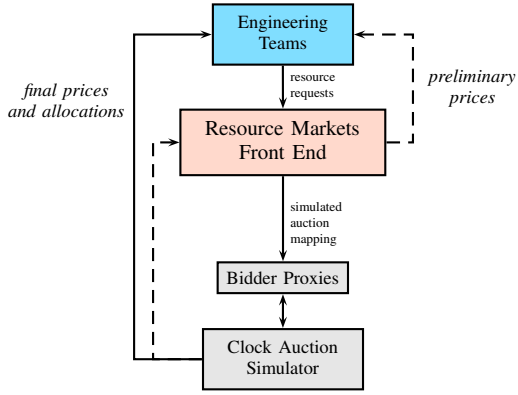
\begin{figure}[t]
\centering \psset{unit=1.8mm}
\begin{pspicture}(25,30)

\scriptsize

\definecolor{lightred}{rgb}{1.0,0.85,0.8}
\definecolor{lightblue}{rgb}{0.5,0.87,1.0}
\definecolor{lightlightgray}{gray}{0.9}

\rput(15,28){\rnode{T}{\psframebox[fillstyle=solid,fillcolor=lightblue]{\begin{tabular}{c}Engineering\\Teams\end{tabular}}}}

\footnotesize

\rput(15,20){\rnode{F}{\psframebox[fillstyle=solid,fillcolor=lightred]{\begin{tabular}{c}Resource Markets\\Front End\end{tabular}}}}

\scriptsize

\rput(15,10){\rnode{P}{\psframebox[fillstyle=solid,fillcolor=lightlightgray]{Bidder Proxies}}}
\rput(15,4){\rnode{C}{\psframebox[fillstyle=solid,fillcolor=lightlightgray]{\begin{tabular}{c}Clock Auction\\Simulator\end{tabular}}}}

\rput(-1,23){\begin{tabular}{c}\emph{final prices}\\\emph{and allocations}\end{tabular}}

\rput(29,24){\begin{tabular}{c}\emph{preliminary}\\\emph{prices}\end{tabular}}

\tiny
\rput[l](14.4,24.2){\begin{tabular}{l}resource\\requests\end{tabular}}

\rput[l](14.4,14){\begin{tabular}{l}simulated\\auction\\mapping\end{tabular}}

\scriptsize

\ncline{->}{T}{F}
\ncline{->}{F}{P}
\ncline{<->}{P}{C}

\ncbar[angle=-180,armA=5]{->}{C}{T}
\ncbar[angle=-180,armA=2.5,linestyle=dashed]{->}{C}{F}
\ncbar[angle=0,linestyle=dashed]{->}{F}{T}

\normalsize

\end{pspicture}

\caption{Bidder interaction with market and clock auction.}
\label{fig:bidderdiagram}
\end{figure}

Given these bids, the trading platform then maps these into a simulated clock auction of the form discussed previously. Each engineering team generally corresponds to one ``user'' in the latter, although teams may be mapped into several, independent ``users'' if they are entering unlinked requests for resources across multiple clusters. In addition, the company itself may be mapped into clock auction participants if it is willing to buy or sell resources in one or more of the pools.

The parameters of this environment are fed into a separate clock auction simulator. The latter, written in Python, runs the clock auction algorithm to determine winning prices and allocations for the (simulated) users. This mapping, simulation, and price update process is run at periodic intervals during the bid collection phase; the preliminary, updated settlement prices are displayed on the market front end, as shown in Figure \ref{fig:bidderdiagram} above. At the conclusion of this phase, one last simulation is run. The results of this are used to determine the final, binding market prices and engineering team allocations.

\subsection{Trading Activity}
\begin{figure}[t]
\centering
\includegraphics[angle=-90,width=8cm]{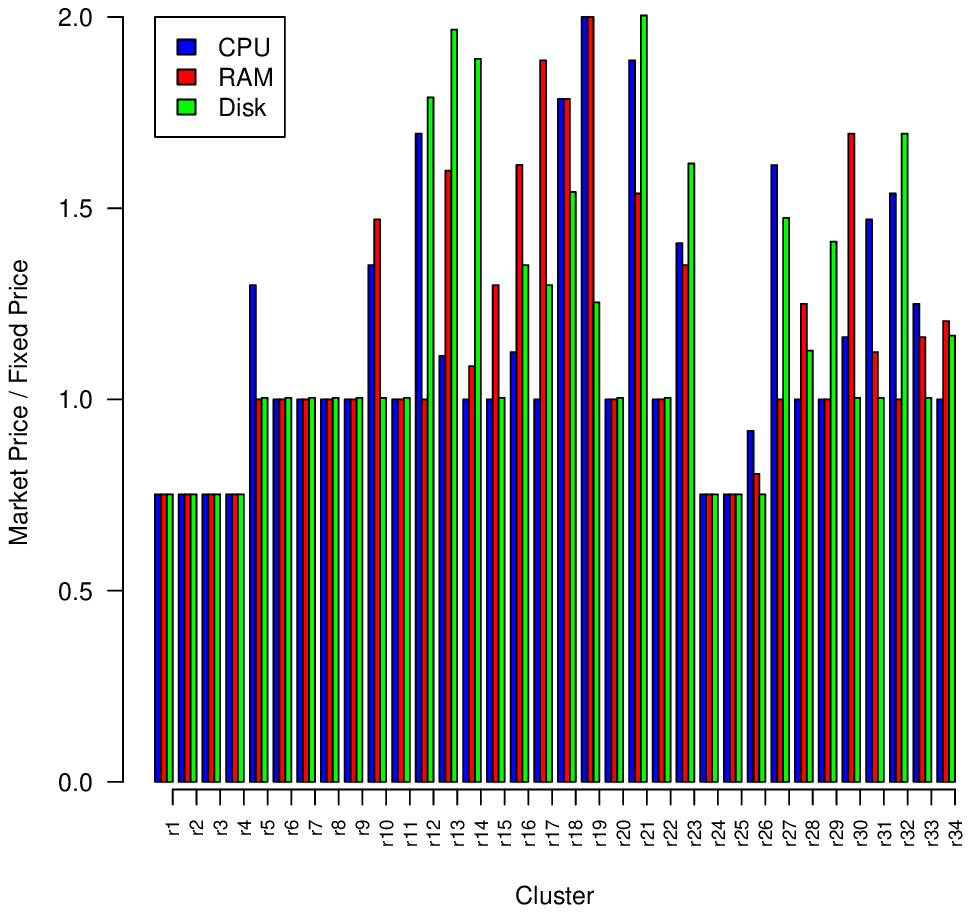}
\caption{Change in resource prices after auction.}
\label{fig:finalprices}
\end{figure}

\begin{figure}[t]
\centering
\includegraphics[angle=-90,width=8cm]{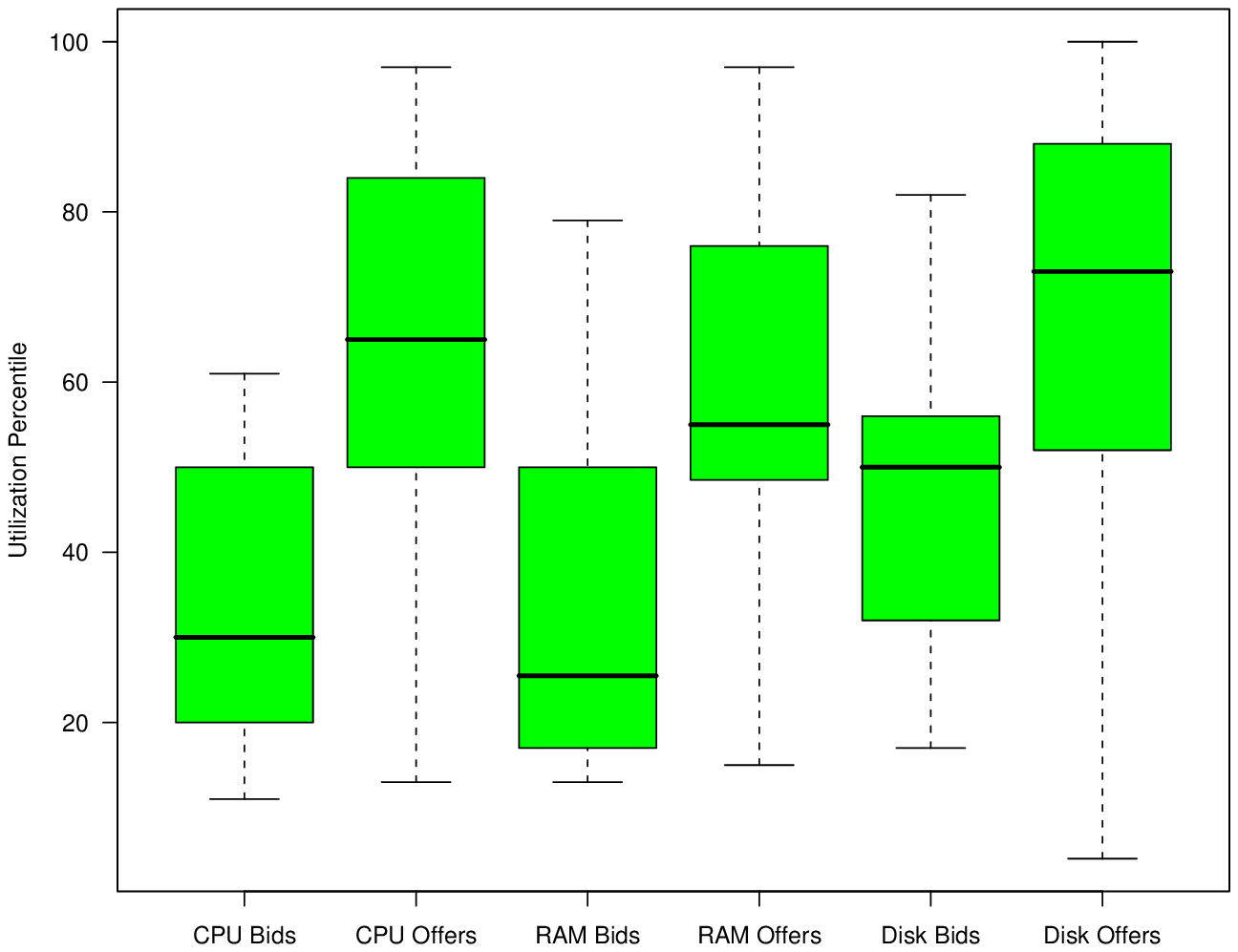}
\caption{Utilization percentiles of resources in settled transactions.}
\label{fig:utilizationboxplot}
\end{figure}

At the time of this writing we have run six, experimental auctions
over the course of several months. As desired, we have seen excess
demand raise the price of resources which were previously
oversubscribed and seen a number of groups move to less crowded clusters.
The plot in Figure \ref{fig:finalprices} shows the ultimate
settlement prices for one of our first auctions as a ratio over the
former fixed price that was in place before the market economy.  The
boxplot in Figure \ref{fig:utilizationboxplot} shows the utilization
percentile of settled trades in the auction broken down by bids and
offers in three resource dimensions.  This plot shows that most bids
were for resources in underutilized clusters and most offers were for
resources in overutilized clusters, which was the behavior strongly
encouraged by the utilization-weighted reserve prices used to start
the clock auction.  It is also interesting to note the significant number of outliers, each representing resource needs for teams  willing to pay a large premium.

It is worth noting that in those clusters with the highest market
prices for resources we saw a number of large teams offer resources on
the market to take advantage of the higher prices and move to less
congested clusters.  We also saw other teams that were willing to
pay a significant price premium to continue growing in congested
clusters even though resources were available at much lower cost
elsewhere.  This discrepancy shows the different premiums teams placed
on relocation.  There is an engineering cost to reconfiguring
applications for different resource pools and the market economy
allows teams to act on those costs autonomously. Without this market, the company would be forced to make centralized policy choices about the best placement of teams with imperfect knowledge of the engineering tradeoffs required.

\subsection{Bidder Behavior}
As the internal market economy has evolved we have noticed a number of
distinct changes in bidder behavior. As users become more familiar
with the market prices we have seen the reserve prices associated with
bids move from closely tracking the former fixed price values to
values much closer to the dynamic market prices. In particular, we
define the premium between the bid price and the ultimate settled
price as

\begin{equation}\label{eq:behavior}
\gamma_u = \frac{\left|\pi_u - x_u^{\textrm{T}} p_u\right|}{x_u^{\textrm{T}}p_u}
\end{equation}

\noindent for each winning user, $u \in \mathbb{W}$.

\begin{table}[t]
\centering
\caption{Bid premium statistics.}
\noindent\begin{tabular}{c|c|c|c}
\hline \hline
Auction & Median of $\gamma_u$ & Mean of $\gamma_u$ & \% Settled\\
\hline
1 & 0.0092 & 0.0614 & 58.9\% \\

2 & 0.0025 & 0.2078 & 88.2\% \\

3 & 0.0009 & 0.0202 & 50.0\% \\
\hline \hline
\end{tabular}
\label{tab:behavior}
\end{table}

Table \ref{tab:behavior} shows the mean and median of this premium
for all bids in the last three auctions.  The fourth column shows the
percentage of trades that were ultimately settled in that auction.  In the earlier auctions bid prices were at times wildly divergent, but the median has decreased significantly
over time.  It is worth pointing out that the mean has been more
variable as in some auctions a number of sellers will enter very low
prices confident that there will be ample competition and that the
final market price will be fair.  Similarly, some bidders in earlier
auctions would enter arbitrarily low bids in the expectation that
these trades would be settled due to lack of competition and excess
Google supply without reserve prices.

Another change in bidder behavior we have observed is an increasing
sophistication towards arbitrage opportunities.  As the market price
differential between resources increases there have been greater
opportunities for teams to profit from one auction to the next.  This
behavior has led to the design of a more robust internal budgeting and
provisioning process inside Google to encourage sharing, mobility, and
thrift by internal teams, and to discourage hoarding and
overestimating. Future research will study these behavioral issues in more detail.

\section{Conclusion}
In this paper, we have thus proposed a framework for allocating and pricing resources in a grid-like environment. This framework employs a market economy with prices adjusted in periodic clock auctions. We have implemented a pilot allocation system within Google based on these ideas. Our preliminary experiments have resulted in significant improvements in overall utilization; users were induced to make their services more mobile, to make disk/memory/network tradeoffs as appropriate in different clusters, and to fully utilize each resource dimension, among other desirable outcomes. In addition, these auctions have resulted in clear price signals, information that  the company and its engineering teams can take advantage of for more efficient future provisioning.

Future research will expand on several dimensions of our work here. On the theoretical side, we intend to more deeply explore existence, convergence, efficiency, and other properties of our clock pricing algorithm. We are also studying some alternative algorithms which explicitly optimize an operator-specified function. On the implementation side, we will expand our market-based allocation system while refining our reserve pricing strategies and overall user experience. These experiments will allow us to better understand user behavior and also to more fully evaluate the utility of employing ``market economies'' for these heterogeneous, large-scale computing environments.

\section*{Acknowledgment}

% Not vfill before pagebreak, it starts new paragraph.
Many others have made contributions to Google's evolving market
economy.  We would like to first acknowledge the work of Srikanth
Rajagopalan in the design and implementation of the market inside
Google as described in the experimental results section.  We would
also like to thank Hal Varian and Cos Nicolaou for early design input
about building a market economy for internal allocation of computing
resources. Our experimental results would not have been possible
\balance
without the help of many internal teams which helped us integrate our
experimental system into the existing tools that engineers use for
provisioning, budgeting, and reserving resources inside Google.
Finally, we are grateful to the early participants in our resource
market, who offered useful practical feedback about the platform.  We
would particularly like to acknowledge the help of Girish Baliga, 
Andrew Barkett, Dan Birken, Hal Burch, Geoffrey Gowan, Vlad Grama,
Alex (Sang) Le, Andrew McLaughlin, and Adam Rogoyski.

\bibliography{IEEEabrv,grid,refs}
\vfill
\end{document}